\documentstyle[prd,aps,tighten,floats,epsfig]{revtex}

\newcommand{\beq}{\begin{equation}}
\newcommand{\eeq}{\end{equation}}
\newcommand{\bear}{\begin{eqnarray}}
\newcommand{\ear}{\end{eqnarray}}
\newcommand{\earn}{\nonumber \end{eqnarray}}
\newcommand{\n}{\label}

\newcommand{\nn}{\nonumber \\}
\newcommand{\Ref}[1]{(\ref{#1})}

\newcommand{\Tmn}{\langle T_{\mu\nu}\rangle}
\newcommand{\ind}[1]{\mbox{\tiny{#1}}}
\newcommand{\calO}{{\cal O}}

\begin{document}
\title{${\langle\varphi^2\rangle}$ for a scalar field in 2D black holes: a new uniform
approximation}

\author{V. Frolov,${}^{(a)}$\thanks{Electronic address:
frolov@phys.ualberta.ca}
S. V. Sushkov,${}^{(b)}$\thanks{Electronic address: sushkov@kspu.kcn.ru}
and
A. Zelnikov${}^{(a,c)}$\thanks{Electronic address:
zelnikov@phys.ualberta.ca}
}

\address{$^{(a)}$Theoretical Physics Institute, Department of Physics,
University of Alberta,
Edmonton, AB, Canada T6G 2J1\\
$^{(b)}$Department of Mathematics, Kazan State Pedagogical University,
Mezhlauk st. 1, Kazan 420021, Russia\\
$^{(c)}$P.N.~Lebedev Physics Institute,
Leninsky pr. 53, Moscow  117924, Russia}

\maketitle

\begin{abstract} We study nonconformal quantum scalar fields
and averages of their local observables (such as
${\langle\varphi^2\rangle}^{\ind{ren}}$ and $ \Tmn^{\ind{ren}}$)
in a spacetime of a 2-dimensional black hole. In order to get an
analytical approximation for these expressions the WKB
approximation is often used. We demonstrate that at the horizon
WKB approximation is violated for a nonconformal field, that is
when the field mass or/and the parameter of non-minimal coupling
do not vanish. We propose a new "uniform approximation" which
solves this problem. We use this approximation to obtain an
improved analytical approximation for
${\langle\varphi^2\rangle}^{\ind{ren}}$ in the 2-dimensional black
hole geometry. We compare the obtained results with
numerical calculations.
\end{abstract}

\section{Introduction}
Calculation of local observables which characterize vacuum
polarization in the gravitational field of a black hole is an
old problem. Main motivation is connected with study of back
reaction effects and constructing a self-consistent model of an
evaporating black hole. A lot of information was obtained
concerning properties of fluctuations ${\langle\varphi^2\rangle}^{\ind{ren}}$  and
vacuum stress-energy tensor $\Tmn^{\ind{ren}}$ in Schwarzschild
and Reissner-Nordstrom geometries. A superscript ${\ind{ren}}$
indicates that we are dealing with physical finite quantities
after a renormalization has been done. Combination of numerical
calculations \cite{HowardCand:84,Howard:84,Ande:90,AnHiSa:95} and
different
analytical approximations
\cite{AnHiSa:95,Cand:80,Page:82,BO:85,BOP:86,FZ:87,FrSuZe:00} usually
gives
quite good results. Nevertheless, there still remains
one problem. Analytical `approximation' does not work in the
vicinity of a black hole horizon if the spacetime is {\em not
Ricci flat} and the quantum field is {\em not conformally
invariant}. It gives logarithmically divergent result near the
horizon in the Hartle-Hawking state, where one expects finite and
smooth behavior. In four dimensions the general structure of this
divergence reads (see e.g. \cite{FrSuZe:00})
\beq \n{1.1}
{\langle\varphi^2\rangle}^{\ind{ren}}\sim a_1\, \ln (|g_{tt}|)\, ,\hspace{0.5cm}
\Tmn^{\ind{ren}}\sim {\delta \int d^4v\, a_2 \over \delta g^{\mu\nu}}\, \ln
(|g_{tt}|)\, .
\eeq
Analogously, in two dimensions it is
\beq
{\langle\varphi^2\rangle}^{\ind{ren}}\sim \, \ln (|g_{tt}|)\, ,\hspace{0.5cm}
\Tmn^{\ind{ren}}\sim {\delta \int d^4v\, a_1 \over \delta g^{\mu\nu}}\, \ln
(|g_{tt}|)\, .
\eeq
Here $a_1$ and $a_2$ are Schwinger-DeWitt coefficients.
For massive scalar fields
\begin{eqnarray}\n{1.2}
a_1 &=& [({1\over 6}-\xi)\, {\cal R}-\mu^2]\, ,\\
a_2 &=& {1 \over 180}\left(
    {\cal R}_{\mu\nu\alpha\beta} {\cal R}^{\mu\nu\alpha\beta}
    - {\cal R}_{\mu\nu} {\cal R}^{\mu\nu}
    + \Box {\cal R} \right)
    +{ 1\over 2}\left[({1\over 6}-\xi)\, {\cal R}-\mu^2\right]^2
    + {1 \over 6}~({1\over 6}-\xi) \Box{\cal R}~,
\end{eqnarray}
where $\mu$ is the mass of the field and $\xi$ is a parameter of
non-minimal coupling with the curvature ${\cal R}$.

In order to obtain the analytical approximation  Anderson,
Hiscock, and Samuel \cite{AnHiSa:95} used WKB approximation for
solutions of radial equations. Recently it was demonstrated that
the breakdown of the analytical approximation (see discussion in
Appendix \ref{A}) is directly related to the breakdown of WKB
approximation for so called zero modes, that is zero-frequency
solutions \cite{KoNaTo:00,Balbinot:01}. By using improved
approximation for zero modes it is possible to correct the
analytical approximation near the horizon and to make all
calculated quantities finite. The concrete methods proposed in
\cite{KoNaTo:00,Balbinot:01} are based on the approximations to
zero modes valid only in the vicinity of the horizon.

The aim of this paper is to develop a new uniform approximation
scheme for calculations of zero-modes contribution to
${\langle\varphi^2\rangle}^{\ind{ren}}$  and $\Tmn^{\ind{ren}}$. Although the method is
applicable in arbitrary dimensions, here, for simplicity, we
restrict ourselves by studying ${\langle\varphi^2\rangle}^{\ind{ren}}$ near
two-dimensional black holes.  We consider different interesting
examples of 2D black hole metrics, and keep the mass of a scalar
field $m$ and the parameter of non-minimal coupling $\xi$ arbitrary.

\section{Green's function}
Consider a massive quantum scalar field with an arbitrary curvature
coupling in a spacetime of the 2-dimensional black hole. The
metric of the static 2-dimensional black hole can be written in
the form
\beq \n{2.1}
dS^2 = -F d t^2 +{dr^2\over F}\, ,
\eeq
where the function $F(r)$ should possess the following properties:
(i) $F(r)$ vanishes at the horizon $r=r_0$ and (ii) $F(r)$ tends
to $1$ at $r\to\infty$. It is convenient to rewrite the metric
(\ref{2.1}) in the dimensionless form as follows
\beq\n{2.4}
ds^2=r_0^{-2}\, dS^2 = -f d\tilde t^2 +{dx^2\over f}\,
,\hspace{1cm}f(x)=F(r)\, ,
\eeq
where $\tilde t=t/r_0$ and $x=(r-r_0)/r_0$, so that the horizon is
located at $x=0$. We shall also use the dimensionless mass $m=\mu r_0$
and curvature $R={\cal R} r_0^2$.
By making Wick's rotation $\tilde t\to i\tau$ in the metric
(\ref{2.4}) one gets the Euclidean metric of the form
\beq \n{euclid}
ds_E^2 =f\, d\tau^2 +{dx^2\over f}\, .
\eeq
It is well known that the property of periodicity of Green functions
in the Euclidean time coordinate $\tau$
with the period $\beta=T^{-1}$, corresponds to quantum systems at
finite temperature $T$.
We consider the case of nonzero temperature $T$ and nonzero surface
gravity $\kappa=f'_x/2\big|_{x=0}$.

The Euclidean Green function $G_E(X,X')$ is a solution of the equation
\beq\n{n2.10}
\left[ \Box_E -m^2-\xi\,R\right]\,
G_E(x,\tau;x',\tau')=-\delta(x,\tau;x',\tau')\,
\eeq
which in the metric (\ref{euclid}) has the following form
\beq\n{n2.11}
\left[f^{-1}{\partial^2\over \partial\tau^2} + {\partial \over
\partial x}\left(f {\partial \over \partial x}\right) -m^2 -\xi
R\right] G_E(x,\tau;x',\tau')=-\delta(\tau-\tau')\, \delta(x-x')
\, .
\eeq
This equation allows a separation of variables, so that we can write
\beq\n{n2.12}
G_E(X,X')= T{\cal G}_0(x,x') +2T\sum_{n=1}^{\infty} \,
\cos\left(\omega_n(\tau -\tau')\right)\, {\cal G}_n(x,x'),
\eeq
where $\omega_n=2\pi n T$ and ${\cal G}_n$ are ``radial'' Green
functions which are the solutions of the 1-dimensional
problem
\beq \n{n2.13}
\left[{d\over dx}\left(f{d\over dx}\right) -
\left(\frac{\omega_n^2}{f}+m^2+\xi R\right)\right]{\cal G}_n(x,x')
= -\delta(x-x')\, .
\eeq
The Green functions ${\cal G}_n$ can be written in the form
\beq\n{Gn}
{\cal G}_n(x,x')=s^{(2)}_n(x^<) s^{(1)}_n(x^>)~,
\eeq
where $x^>=\max(x,x')$ and $x^<=\min(x,x')$. Here the radial modes
$s^{(1)}_n$ and $s^{(2)}_n$ are two linear independent solutions
of the homogeneous equation of motion
\beq \n{eqmo1}
\left[{d\over dx}\left(f{d\over dx}\right) -
\left({\omega_n^2\over f}+m^2+\xi R\right)\right] s(x) = 0~.
\eeq
We choose $s^{(1)}_n$ to be regular at infinity and $s^{(2)}_n$ to be
regular at the horizon. The solutions
$s^{(1)}_n$ and $s^{(2)}_n$ are normalized by
the condition
\beq\n{normcond}
s^{(1)}_n{ds^{(2)}_n\over dx}-s^{(2)}_n{ds^{(1)}_n\over
dx}=-\frac1{f}~.
\eeq

The coincidence limit of the Euclidean Green function gives
the unrenormalized expression for $\langle\varphi^2\rangle$, which is
ultraviolet divergent,
\beq\n{phiunren}
\langle\varphi^2(x)\rangle_{\rm unren}=G_E(x,\tau;x,\tau')=T{\cal
G}_0(x,x) +2T\sum_{n=1}^{\infty} \, \cos(\omega_n(\tau -\tau'))\,
{\cal G}_n(x,x).
\eeq
(For convenience the points are split in time direction only,
so $x'=x$.)
Subtraction of ultraviolet divergencies gives the
renormalized ${\langle\varphi^2\rangle}$
\beq
{\langle\varphi^2\rangle}_{\rm ren}=\lim_{\tau'\to\tau} \left({\langle\varphi^2\rangle}_{\rm
unren}-{\langle\varphi^2\rangle}_{\rm DS}\right)~,
\eeq
where ${\langle\varphi^2\rangle}_{\rm DS}$ is the DeWitt-Schwinger counterterm which
in two dimensions has the following form
\beq\n{phiDS}
{\langle\varphi^2\rangle}_{\rm
DS}=-\frac1{4\pi}(\ln\frac{\mu^2\sigma}{2}+2C)~.
\eeq
Here $\sigma$ is equal to one half the square of the distance
between the points $x$ and $x'$ along the shortest geodesic
connecting them, $C$ is the Euler's constant. For a massive fields
the constant $\mu$ is
equal to its mass. For a
massless field it is an arbitrary mass scale parameter. A particular
choice of the value of $\mu$ corresponds to a finite
renormalization of the coefficients of terms in the gravitational
Lagrangian.

The stress-energy tensor can be obtained by applying a certain
differential operator to the Green function:
\beq
\Tmn=\lim_{X'\to X}D_{\mu\nu}G_E(X,X').
\eeq

In order to calculate ${\langle\varphi^2\rangle}$ and $\Tmn$ 
one needs to solve the radial equation of motion \Ref{eqmo1}
and to find the radial Green functions ${\cal G}_n(x,x)$. 
In a general case  it is
impossible to find exact solutions of this equation, and so one
oftes uses various approximate methods. In case the scalar
field has a nonzero mass,\footnote{Generally speaking, the mass
should be large enough. If the mass is small or vanishes the WKB
approximation cannot be applied for modes with small numbers $n$,
including the zero mode, $n=0$.} one of the standard approaches is
the WKB approximation. For example, recently Anderson, Hiscock and
Samuel have derived the analytic approximation for ${\langle\varphi^2\rangle}$ and
$\Tmn$ for a scalar field in the general static spherically
symmetric spacetime using the WKB expansion for radial modes
\cite{AnHiSa:95}. However, the serious obstacle for applying this method
in black hole spacetimes is that the WKB approximation even in
case $m>0$ breaks down at the horizon for low-frequency modes.

In the section \ref{secWKB} we discuss this problem in more detail.

\section{Exactly solvable models} 

Let us consider the radial equation of motion \Ref{eqmo1}. Let us
introduce new variables
\beq\label{changecoord}
\frac{dx}{f(x)}=\frac{dy}{y}, \quad s(x)=\frac{u(y)}{y^{1/2}},
\eeq
and rewrite this equation in the form
\beq\label{eqmo2}
\left[\frac{d^2}{dy^2}-\left(\frac{\omega_n^2-\frac14}{y^2}
+\frac{(m^2+\xi R){\cal F}}{y^2}\right)\right] u(y)=0~.
\eeq
Here ${\cal F}(y)=f(x)$  and
\beq\n{R}
R=-\frac{y^2 {\cal F}''}{{\cal F}^2}-\frac{y {\cal F}'}{{\cal F}^2}
  +\frac{y^2 {\cal F}'^2}{{\cal F}^3}~.
\eeq
The prime means the derivative with respect to $y$.

First of all, let us discuss cases when the equation
\Ref{eqmo2} can be solved exactly.

\subsection{Conformally invariant scalar field}

Consider the 2d conformally invariant scalar field, i.e.,
assume $m=0$ and $\xi=0$. In this case the equation \Ref{eqmo2}
takes the form
\beq
\left[\frac{d^2}{d y^2}-\frac{\omega_n^2-\frac14}{y^2}\right]u=0.
\eeq
Two independent solutions of this equation are
\bear
&&u_{n=0}^{(1)}=y^{1/2},\quad u_{n=0}^{(2)}=-y^{1/2}\,\ln y,\\
&&u_{n>0}^{(1)}=\frac1{\sqrt{2\omega_n}}\,y^{\omega_n+1/2},\quad
u_{n>0}^{(2)}=\frac1{\sqrt{2\omega_n}}\,y^{-\omega_n+1/2}.
\ear


\subsection{2D dilatonic black hole}

Consider a scalar field in the 2D dilatonic black hole spacetime
with (see \cite{FZ:01})
\beq
f(x)=1-e^{-x}.
\eeq
Using the coordinate $y$ we find
\beq\n{fdilaton}
{\cal F}(y)={y\over 1+y},
\eeq
and
\beq
R=\frac{1}{1+y}~.
\eeq
The equation \Ref{eqmo2} now reads
\beq\label{hypereq}
\frac{d^2u}{dy^2}-\left(\frac{\omega_n^2-\frac14}{y^2}
+\frac{m^2}{y (1+y)}+\frac{\xi }{y(1+y)^2}\right) u=0.
\eeq
This equation can be solved in terms of hypergeometric functions
\cite{FZ:01}. Its two independent solutions are
\bear
u^{(1)}_n&=&y^{1/2-\Omega_n}\left(\frac{y}{1+y}\right)^{\omega_n+\Omega_n}
F(a,b; 2\omega_n+1; \frac{y}{1+y}), \nn \n{FZ}
u^{(2)}_n&=&y^{1/2-\Omega_n}\left(\frac{y}{1+y}\right)^{\omega_n+\Omega_n}F(a,b;
2\Omega_n+1; \frac1{1+y}),
\ear
The Wronskian of these solutions is
\bear
W[u^{(1)}_n,u^{(2)}_n]\equiv
W_n&=&u^{(1)}_n\frac{du^{(2)}_n}{dy}-\frac{du^{(2)}_n}{dy}u^{(1)}_n
=\frac{\Gamma(2\omega_n+1)\Gamma(2\Omega_n+1)}{\Gamma(a)\Gamma(b)},
\ear
where
$$
\Omega_n=\sqrt{\omega_n^2+m^2},
$$
and
$$
a=\frac12+\omega_n+\Omega_n+\frac12\sqrt{1-4\xi}~, \quad
b=\frac12+\omega_n+\Omega_n-\frac12\sqrt{1-4\xi}~.
$$

\section{The WKB approximation for radial modes}\label{secWKB}

In the cases when the analytical solution is not known 
WKB approximation is often used.
Let us write the radial modes $s^{(1)}_n$ and $s^{(2)}_n$ in
the WKB form
\beq\n{WKBmode1}
s^{(1)}_n(x)={1\over \sqrt{2\Omega(x)}}
\exp\left(\int^x{\Omega(x')\over f(x')}dx'\right),
\eeq
\beq\n{WKBmode2}
s^{(2)}_n(x)={1\over \sqrt{2\Omega(x)}}
\exp\left(-\int^x{\Omega(x')\over f(x')}dx'\right),
\eeq
where $\Omega(x)$ is a new unknown function. Note that these
solutions are properly normalized to satisfy the
condition \Ref{normcond}. Substituting \Ref{WKBmode1},
\Ref{WKBmode2} into \Ref{eqmo1} gives the equation for
$\Omega(x)$:
\beq\label{Omega2}
\Omega^2=\omega_n^2+m^2 f+\xi R f+\frac12{f^2\Omega''\over\Omega}
+\frac12{ff'\Omega'\over\Omega}
-\frac34{f^2\Omega'^2\over\Omega^2}.
\eeq
This equation can be solved iteratively with the zeroth-order solution
chosen as
\beq
\Omega^{(0)}=\Omega_0=\left[\omega_n^2+m^2 f\right]^{1/2}.
\eeq
The first-order solution is
\beq\n{2order}
\Omega^{(1)}=\Omega_0+\Omega_1=\Omega_0+\frac{\xi Rf}{2\Omega_0}
+\frac14\frac{f^2\Omega_0''}{\Omega_0^2}
+\frac14\frac{ff'\Omega_0'}{\Omega_0^2}
-\frac38\frac{f^2{\Omega_0'}^2}{\Omega_0^3}~.
\eeq
Analogously, the $k$'th-order solution can be found
\beq\n{2k}
\Omega^{(k)}=\Omega_0+\Omega_1+\dots+\Omega_{k},
\eeq
The WKB approximation is based on an assumption that
\beq\n{WKBrel}
\Omega_0\gg\Omega_1\gg\dots\gg\Omega_{k}.
\eeq
This ensures the convergence of series
$\sum_{k=0}^{\infty}\Omega_{k}$ and allows one to break the series
in order to construct an approximate solution for $\Omega$ of the
form \Ref{2k}. Substituting $\Omega^{(k)}$ into \Ref{WKBmode1},
\Ref{WKBmode2} gives the WKB approximation of $k$'th order for
the radial modes $s^{(1,2)}_n$. By and substituting $s^{(1,2){\rm
WKB}}_n$ into \Ref{Gn} and then into \Ref{phiunren} one obtains the WKB
approximation for $\langle\varphi^2\rangle_{\rm unren}$:
\beq\n{phiWKB}
\langle\varphi^2(x)\rangle_{\rm unren,WKB}= T{\cal G}_0^{\rm
WKB}(x,x) +2T\sum_{n=1}^{\infty} \, \cos(\omega_n(\tau -\tau'))\,
{\cal G}_n^{\rm WKB}(x,x),
\eeq
with
\beq\label{Gn}
{\cal G}_n^{\rm WKB}(x,x)=s_n^{(1)\rm WKB}(x)s_n^{(2)\rm WKB}(x)~.
\eeq

Let us discuss the validity of this approximation for zero-modes.
i.e., when $n=0$. 
For $n=0$ one has $\Omega_0=mf^{1/2}$, i.e.,
$\Omega_0\sim f^{1/2}$. Then $\Omega_1\sim {f'}^2 f^{-1/2}$ (see
Eq. \Ref{2order}). Note that at the horizon the function $f$ vanishes
while $f'$ and $\kappa$ remain finite. Thus in the vicinity of the horizon 
$\Omega_0\ll\Omega_1$.
Moreover, one can show that near the horizon
inequalities $\Omega_0\ll\Omega_1\ll\dots\ll\Omega_{k}$ are
fulfilled instead of \Ref{WKBrel}. This means that the series
$\sum_{k=0}^{\infty}\Omega_{k}$ diverges, and the WKB expansion
breaks down. Thus, the WKB method can not be used for
constructing ${\cal G}_{n=0}$ near the horizon.

\section{New uniform approximation for the radial modes} \label{ua}

\subsection{General discussion}
In this
section we propose a new approximation which can be used instead of 
the WKB approximation.

We choose the time $T$ normalization so that 
the function ${\cal F}(y)$  tends to $1$ at 
$y\to\infty$. 
We assume also that the surface gravity does not vanish, so that
${\cal F}(y)\sim y$ at $y\to0$. Let us write ${\cal F}(y)$ in the form
\beq\n{genrep}
{\cal F}(y)=\frac{y{\em h}(y)}{c+y}~.
\eeq
Here $c>0$ is a positive constant, and ${\em h}(y)$ is a
positive function
depending on $c$ and having the asymptotics
\bear
{\em h}(y)|_{y\to0}&=&{\em h}_0+\calO(y)~,\hskip 2cm {\em h}_0> 0\nonumber\\
{\em h}(y)|_{y\to\infty}&=&1+\calO(y^{-1})~.
\ear
Note that such the representation
explicitly reflects the asymptotical properties of ${\cal F}(y)$:
\bear
{\cal F}(y)|_{y\to0}&=&{\em h}_0 c^{-1}y+\calO(y^2),\nonumber\\
{\cal F}(y)|_{y\to\infty}&=&1+\calO(y^{-1}).
\ear
The expression \Ref{R} for the scalar curvature now reads
\beq\n{Rphi}
R=\frac{c}{(c+y){\em h}}-(c+y)\frac{{\em h}'}{{\em h}^2}
-y(c+y)\frac{{\em h}''}{{\em h}^2}+y(c+y)\frac{{{\em h}'}^2}{{\em
h}^3}~.
\eeq
The equation \Ref{eqmo2} takes the form
\beq\n{eqmogen}
\frac{d^2 u}{d y^2}-\left(\frac{\omega_n^2-\frac14}{y^2}
+\frac{m^2{\em h}}{y(c+y)}+\frac{c\xi}{y(c+y)^2}+\xi V\right)u=0,
\eeq
where
\beq\n{defV}
V(y)=-\frac{{\em h}''}{{\em h}}-\frac{{\em h}'}{y{\em h}}
+\frac{{\em h}'^2}{{\em h}^2}.
\eeq
In the special case $c=1$ and ${\em h}(y)\equiv 1$ we obtain
the 2D dilatonic black hole with ${\cal F}(y)=y/(1+y)$ and $V=0$ (see
Eqs. \Ref{fdilaton}, \Ref{hypereq}).

In a general case we will fix the constant $c$ by the algebraic
condition
\beq\n{fixc}
{\em h}'(0)=0,
\eeq
in order that to guarantee the regular behavior of $V(y)$ near the
horizon $y=0$. This condition shows that the function ${\em h}(y)$
near $y=0$ has the form:
\beq\n{near0}
{\em h}(y)|_{y\to0}={\em h}_0+\calO(y^2).
\eeq
We rewrite the equation \Ref{eqmogen} as
\beq\n{short}
\frac{d^2 u}{d y^2}-(U_n+\xi V)u=0,
\eeq
where we denote
\beq\n{defU}
U_n(y)=\frac{\omega_n^2-\frac14}{y^2}
+\frac{m^2{\em h}}{y(c+y)}+\frac{c\xi}{y(c+y)^2}.
\eeq
Let us compare two terms $U_n$ and $\xi V$. In case $\xi=0$ 
the term $\xi V$ disappears and the equation \Ref{short} takes the
form
\beq\n{shortxi0}
\frac{d^2 u}{d y^2}-U_n u=0.
\eeq
Now assume that $\xi$ does not vanish. Let us consider separately
asymptotical regions near the horizon ($y\to0$) and far from it
($y\to\infty$). Using the asymptotical form \Ref{near0}
for ${\em h}(y)$ we can find that near the horizon
$V(y)=\calO(y^0)$ and $U_n(y)=\calO(y^{-2})$, and hence in the
limit $y\to0$ the absolute value of the term $\xi V$ is much less
than the one of $U_n$. Far from the horizon 
 ${\em h}(y)=1+\calO(y^{-1})$, then $V(y)=\calO(y^{-3})$ and
$U_n(y)=\calO(y^{-2})$, and so we can again conclude that $|\xi
V|\ll |U_n|$ in the limit $y\to\infty$.

In order to estimate values of the terms $\xi V$ and $U_n$ in the
intermediate region $0<y<\infty$ we consider the particular
configuration: two dimensional ``Schwarzschild'' black hole with
\beq\n{fschw}
f(x)={x\over x+1}.
\eeq
By using the relation ${dx/f(x)}={dy/ y}$ we
find
\beq\n{y2x}
y=xe^x.
\eeq
${\em h}$ as a function of $x$ can be obtained by comparing two
representations, \Ref{genrep} and \Ref{fschw}, for $f$:
$$
{\cal F}(y)={y{\em h} \over c+y}\, ,
$$
so that
\beq\n{phischwarzschild}
{\em h}\big(y(x)\big)={c+xe^x \over (x+1)e^x}.
\eeq
The constant $c$ can be fixed by the condition \Ref{fixc}, which
gives $c=1/2$.
$V$ and $U_n$  calculated for this metric
are shown in the figures 1 and 2. In
the figure 1 graphs for $y^2U_{n=0}$ are given for various values
of $m$ and $\xi$. Note that in case $m<0.5$ the function
$y^2U_{n=0}(y)$ is strictly negative, while for larger masses
$m>0.5$ it can change  sign, so that there is a
point $y_*>0$ where $y_*^2U_{n=0}(y_*)=0$. Hereafter we will
assume that $m<0.5$. As
for the $n>0$ modes one can easily see from Eq. \Ref{defU} that
the function $y^2U_{n>0}(y)$ is positive for arbitrary $m\ge 0$
and $\xi\ge 0$.
\begin{figure}\n{fig1}
\centerline{\includegraphics{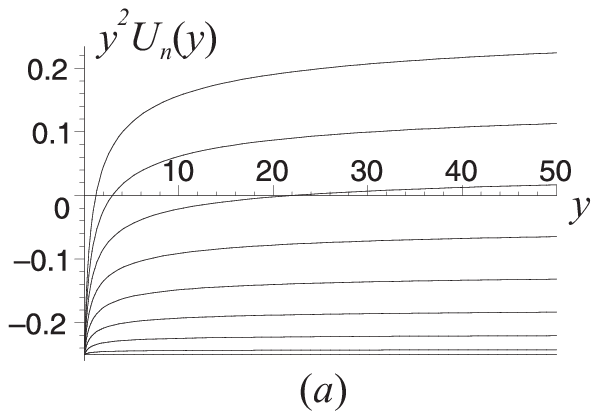}\quad
\includegraphics{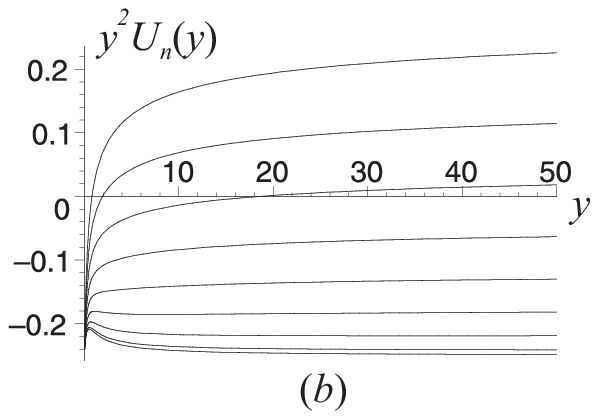}} \caption{The graphs of the function
$y^2U_{n=0}(y)$; (a) $\xi=0$; (b) $\xi=1/6$; the graphs
corresponds to the values $m=0$, $0.1$, $0.2$, $0.3$, $0.4$,
$0.5$, $0.6$, $0.7$, $0.8$ from bottom to top.}
\end{figure}
The graph of $y^2V(y)$ is given in the figure 2(a). Note that the
function $V(y)$ is completely determined by ${\em h}(y)$ and does
not depend on the parameters $m$ and $\xi$. In the figure 2(b) the
ratio $\xi V/U_n$ is shown for $n=0$, $1$, $2$, $m=0.1$, and
$\xi=1/6$.
\begin{figure}\n{fig2}
\centerline{\includegraphics{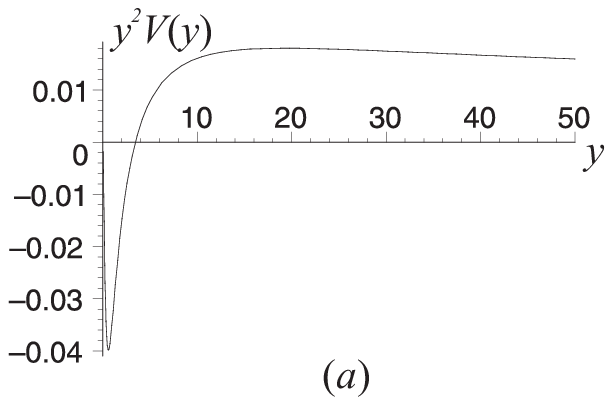}\quad
\includegraphics{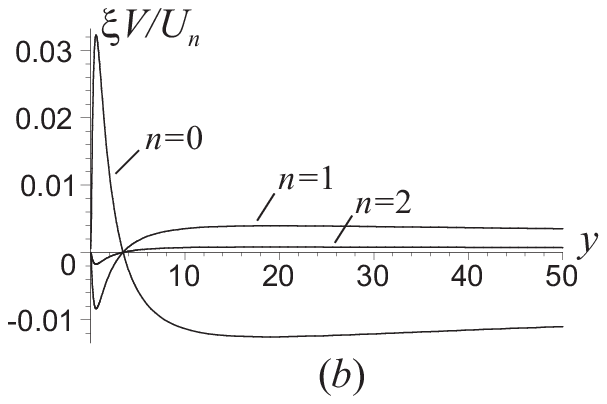}} \caption{{\bf (a)} The graph of the function
$y^2V(y)$. {\bf (b)} The ratio ${\xi V/U_n}$; $n=0,1,2$, $m=0.1$,
$\xi=1/6$; $T=(4\pi)^{-1}$. }
\end{figure}
One has $|\xi V/U_n|\ll 1$. For example, $\max(|\xi
V/U_{n}|)\approx 0.032$, $0.009$, $0.0002$ for $n=0$, $1$, $2$
respectively, so that the higher the number of mode the less the
ratio $|\xi V/U_n|$.

Thus we may conclude that at least for 2D Schwarzschild metric 
the term $\xi V(y)$ is much less than
$U_n(y)$ in the whole range of $y$. Using the smallness of $\xi V$
in comparison with $U_n$ we can neglect this term in the equation
\Ref{eqmogen}.


\subsection{The uniform approximation for the massless field modes}

Consider the massless case, $m=0$, with arbitrary coupling $\xi$.
Neglecting the term $\xi V$ in the equation \Ref{eqmogen} we
obtain
\beq\n{eqmoapp0}
\frac{d^2 u}{d y^2}-\left(\frac{\omega_n^2-\frac14}{y^2}
+\frac{c\xi}{y(c+y)^2}\right)u=0,
\eeq
or, after rescaling $y=c\tilde y$,
\beq
\frac{d^2 u}{d\tilde y^2}-\left(\frac{\omega_n^2-\frac14}{\tilde
y^2} +\frac{\xi}{\tilde y(1+\tilde y)^2}\right)u=0~.
\eeq
Comparing with Eqs. \Ref{hypereq}, \Ref{FZ} we find two
independent solutions of the equation \Ref{eqmoapp0}:
\bear\n{approx0}
u^{(1)}_n&=&y^{1/2+\omega_n}(c+y)^{-2\omega_n} F(a,b; 2\omega_n+1;
\frac{y}{c+y}), \nn
u^{(2)}_n&=&y^{1/2+\omega_n}(c+y)^{-2\omega_n}F(a,b; 2\omega_n+1;
\frac{c}{c+y}),
\ear with
$$
a=\frac12+2\omega_n+\frac12\sqrt{1-4\xi}, \quad
b=\frac12+2\omega_n-\frac12\sqrt{1-4\xi}.
$$
These functions represent an approximate solution of the equation
\Ref{eqmogen} in case $m=0$.


\subsection{The uniform approximation for the massive field modes}

Consider the massive case. Neglecting the term $\xi V$ in
the equation \Ref{eqmogen} gives
\beq\n{eqmoapp}
\frac{d^2 u}{d y^2}-\left(\frac{\omega_n^2-\frac14}{y^2}
+\frac{m^2{\em h}(y)}{y(c+y)}+\frac{c\xi}{y(c+y)^2}\right)u=0.
\eeq
Generally this equation cannot be solved exactly, and so
let us construct an approximate solution. For this aim we study
asymptotical properties of the equation \Ref{eqmoapp}. First
consider the region far from the horizon, $y\to\infty$. Neglecting
terms of order $y^{-3}$ and smaller in the equation \Ref{eqmoapp}
we get
\beq\n{eqnear8}
\left[\frac{d^2}{dy^2}-\frac{\omega_n^2-\frac14+m^2{\em h}}{y^2}
\right]u=0.
\eeq
Introducing in the equation \Ref{eqnear8} a new
variable\footnote{The variable $x_*$ is known as a tortoise
coordinate, $dx/f(x)=dy/y=dx_*$.} by $dx_*=dy/y$ and using
the relation $s(x_*)=y^{-1/2}u(y)$ we obtain
\beq\n{canonical}
\frac{d^2s}{dx_*^2}-\Omega_n^2(x_*)s=0,
\eeq
where
\beq
\Omega_n(x_*)=\sqrt{\omega_n^2+m^2{\em h}}.
\eeq
An approximate solution of the equation \Ref{canonical} can be
written in the well-known WKB form:
\beq
s_n^{(1)}=\frac{1}{\sqrt{2\Omega_n}}\,\exp\left[\int\Omega_n
dx_*\right],\quad
s_n^{(2)}=\frac{1}{\sqrt{2\Omega_n}}\,\exp\left[-\int\Omega_n
dx_*\right].
\eeq
Returning to the variables $y$ and $u_n$ we obtain the WKB
solution of the equation \Ref{eqnear8}:
\beq\n{asymp8}
u_n^{(1)}=\frac{y^{1/2}}{\sqrt{2\Omega_n}}\,\exp\left[\int\Omega_n
\frac{dy}{y}\right],\quad
u_n^{(2)}=\frac{y^{1/2}}{\sqrt{2\Omega_n}}\,\exp\left[-\int\Omega_n
\frac{dy}{y}\right].
\eeq
These functions give an approximate solution of the
equation \Ref{eqmoapp} in the region $y\to\infty$.

Now consider the region near the horizon, $y\to0$. Taking into
account Eq.(\ref{near0}) and
neglecting in the equation \Ref{eqmoapp} terms of order $y^{-1}$
and smaller we get
\beq\n{eqnear0}
\left[\frac{d^2}{dy^2}-\frac{\omega_n^2-\frac14}{y^2} \right]u=0.
\eeq
It is worth noting that the last equation does not contain the
parameters $m$ and $\xi$. This means that any scalar
field with arbitrary mass $m$ and coupling $\xi$ near a horizon
behaves effectively like the conformal scalar field with $m=0$ and
$\xi=0$. Two independent solutions of the equation \Ref{eqnear0}
are
\bear
u_{n=0}^{(1)}&=&y^{1/2},\quad u_{n=0}^{(2)}=-y^{1/2}\ln y\\
u_{n>0}^{(1)}&=&\frac1{\sqrt{2\omega_n}}\,y^{\omega_n+1/2},\quad
u_{n>0}^{(2)}=\frac1{\sqrt{2\omega_n}}\,y^{-\omega_n+1/2}.
\ear
Respectively, for the radial modes
$s_n^{(1,2)}=y^{-1/2}u_n^{(1,2)}$ expressed in the coordinate $y$
[see Eq. \Ref{changecoord}] we obtain
\bear\n{asymp0}
s_{n=0}^{(1)}&=&1,\quad
s_{n=0}^{(2)}=-\ln y\\
s_{n>0}^{(1)}&=&\frac1{\sqrt{2\omega_n}}\,y^{\omega_n},\quad
s_{n>0}^{(2)}=\frac1{\sqrt{2\omega_n}}\,y^{-\omega_n}.
\ear
Note that the modes $s_n^{(1,2)}$ are normalized by the condition
\Ref{normcond} which for the coordinate $y$ reads
\beq\n{normcondy}
s^{(1)}_n{ds^{(2)}_n\over dy}-s^{(2)}_n{ds^{(1)}_n\over
dy}=-\frac1{y}.
\eeq

Thus, we may summarize that a solution of the equation \Ref{eqmoapp}
should possess the asymptotical properties \Ref{asymp8} and
\Ref{asymp0}. We also assumed that terms containing derivatives of
${\em h}$ are small and can be neglected. At last, it should be taken
into account that in case ${\em h}\equiv{\rm const}$ the equation
\Ref{eqmoapp} could be solved analytically with the solutions given
by \Ref{FZ}.\footnote{More exactly speaking, in order to obtain the
solution of the equation \Ref{eqmoapp} in the form \Ref{FZ} in case
${\em h}\equiv{\rm const}$ one has to make rescaling $\tilde
m^2=m^2{\em h}$ and $\tilde y=c^{-1}y$.} Now combining these results
we choose approximate solutions of the equation \Ref{eqmoapp} as
follows:
\bear\n{appsol}
u_n^{(1)}&=&\frac{y^{1/2}}{\sqrt{2\Omega_n}}\,\exp\left[\int\Omega_n
\frac{dy}{y}\right]\Psi_n^{(1)},\nonumber \\
u_n^{(2)}&=&\frac{y^{1/2}}{\sqrt{2\Omega_n}}\,\exp\left[-\int\Omega_n
\frac{dy}{y}\right]\Psi_n^{(2)},
\ear
with
\bear
\Psi_n^{(1)}&=&\frac{2\Omega_n}{W_n}\left(\frac{y}{c+y}\right)^{\omega_n+\Omega_n}
y^{-2\Omega_n}F(a,b;2\omega_n+1;\frac{y}{c+y}),\nonumber \\
\Psi_n^{(2)}&=&\left(\frac{y}{c+y}\right)^{\omega_n+\Omega_n}
F(a,b;2\Omega_n+1;\frac{c}{c+y}).
\ear
It can be shown that the given functions $u_n^{(1,2)}$ possess all
necessary properties. First, substituting the solutions
\Ref{appsol} into the equation \Ref{eqmoapp} one may see that they
obey the equation up to terms containing derivatives of ${\em h}$.
Second, using properties of the hypergeometric functions (see Ref.
\cite{Abram}) one can verify that the solutions \Ref{appsol} have
the asymptotical form \Ref{asymp0} and \Ref{asymp8} at $y\to0$ and
$y\to\infty$, respectively. Moreover, note that the functions
\Ref{appsol} reproduce the exact solution \Ref{FZ} in case
${\em h}\equiv{\rm const}$ or/and $m=0$.

Finally, using the relation $s(x)=y^{-1/2}u(y)$ and the formulas
\Ref{appsol} gives an approximation for the radial modes
$s^{(1,2)}_n.$ It is worth noting that this approximation has been
derived for arbitrary mass $m$ of the scalar field (including
$m=0$) and arbitrary coupling $\xi$. It is also important that the
approximation works correctly both far from and near the horizon.
For this reason we will call it as a {\em uniform}\,
approximation.

\section{Evaluating of ${\langle\varphi^2\rangle}$ in  the uniform approximation}
Using the uniform approximation, we may obtain the
radial Green functions ${\cal G}_n=s^{(1)}_ns^{(2)}_n$ and
construct the approximate expression for
${\langle\varphi^2\rangle}_{\rm unren}$:
\beq\n{phiuniform}
{\langle\varphi^2\rangle}_{\rm unren}= T{\cal G}_0^{\rm ua}(x,x)
+2T\sum_{n=1}^{\infty} \, \cos(2\pi nT\epsilon)\, {\cal G}_n^{\rm
ua}(x,x),
\eeq
where the superscript `ua' is used to denote the uniform
approximation.

We may simplify the resulting approximate expression for
${\langle\varphi^2\rangle}_{\rm unren}$ if we take into account
that the $n>0$ modes are satisfactory described by the WKB
approximation. In this case we will use the uniform approximation
in order to compute ${\cal G}_{n=0}$ only, and use the WKB
approximation in order to compute the other radial Green functions
${\cal G}_{n>0}$, so that
\beq
{\langle\varphi^2\rangle}_{\rm unren}= T{\cal G}_0^{\rm ua}(x,x)
+2T\sum_{n=1}^{\infty} \, \cos(2\pi nT\epsilon)\, {\cal G}_n^{\rm
WKB}(x,x).
\eeq
The corresponding calculations for $\langle T^\mu_\nu \rangle$
would require to use the uniform approximation for $n=1$ as well.
Subtracting ${\langle\varphi^2\rangle}_{\rm DS}$ from
${\langle\varphi^2\rangle}_{\rm unren}$ and taking the limit
$\epsilon\to0$ gives the renormalized expression for
${\langle\varphi^2\rangle}$:
\beq
{\langle\varphi^2\rangle}_{\rm ren}={\langle\varphi^2\rangle}_0+{\langle\varphi^2\rangle}_1,
\eeq
where we denote
\beq
{\langle\varphi^2\rangle}_0=T{\cal G}_0^{\rm ua}(x,x),
\eeq
and
\beq\label{phi1^2}
{\langle\varphi^2\rangle}_1=\lim_{\epsilon\to0}\left[2T\sum_{n=1}^{\infty} \,
\cos(2\pi nT\epsilon)\, {\cal G}_n^{\rm WKB}(x,x)-{\langle\varphi^2\rangle}_{\rm DS
}\right].
\eeq
Using the uniform approximation \Ref{appsol} for the radial modes
$s=y^{-1/2}u$ and taking into account that
$$
\omega_0=0,\quad \Omega_0=m\sqrt{{\em h}}, \quad
W_0=\frac{\Gamma(2m\sqrt{{\em h}}+1)}{\Gamma(a)\Gamma(b)},
$$
$$
a=\frac12+m\sqrt{{\em h}}+\frac12\sqrt{1-4\xi},\quad
b=\frac12+m\sqrt{{\em h}}-\frac12\sqrt{1-4\xi},
$$
we can write ${\langle\varphi^2\rangle}_0=T{\cal G}_0^{\rm ua}=Ts^{(1)}_0s^{(2)}_0$ as
follows:
\bear\n{phisq0}
{\langle\varphi^2\rangle}_0&=&\frac{T}{2\Omega_0}\Psi^{(1)}_0\Psi^{(2)}_0 \nonumber \\
&=&\frac{T\,\Gamma(a)\Gamma(b)}{\Gamma(2m\sqrt{{\em
h}(y)}+1)}\left(\frac{1}{c+y}\right)^{2m\sqrt{{\em h}(y)}}
F\left(a,b;1;\frac{y}{c+y}\right)F\left(a,b;2m\sqrt{{\em
h}(y)}+1;\frac{c}{c+y}\right).
\ear
For a given spacetime the
function ${\em h}$ and the parameter  $c$ are known. Hence the
expression \Ref{phisq0} represents an analytical formula
describing an approximation for ${\langle\varphi^2\rangle}_0$.

To evaluate ${\langle\varphi^2\rangle}_1$ one may use the
well-elaborated summation procedure dealing with the WKB
approximation (see, e.g., \cite{Ande:90,AnHiSa:95}, and also
\cite{Su}). Combining
Eqs.(\ref{Omega2},\ref{Gn},\ref{phi1^2},\ref{sum_n}), we obtain
\begin{eqnarray}
{\langle\varphi^2\rangle}_1&=&\frac{1}{4\pi}\left[\ln\left(\frac{\mu^2f}{2(2\pi
T)^2}\right)+2C\right]+{1\over 2\pi}\,S_0
\nonumber\\
&&+T\left[-\frac12\xi RfS_1-\frac18m^2(f^2f''+f{f'}^2)S_2
+\frac5{32}m^4f^2{f'}^2S_3\right]
\end{eqnarray}
where
\beq
S_0=\sum_{n=1}^\infty \left[{1\over\sqrt{n^2+{m^2f\over(2\pi
T)^2}}}-{1\over n}\right],
\eeq
and
\beq
S_k=\sum_{n=1}^\infty\frac1{\left[(2\pi
Tn)^2+m^2f\right]^{k+1/2}}, \quad k=1,2,3,\dots .
\eeq
All the sums $S_j$ in this formula converge everywhere and $S_0$
tends to zero at the horizon ($f\rightarrow 0$).
The logarithmic term combined with
${\langle\varphi^2\rangle}_0$, which also logarithmically diverges
at the horizon, leads to $(T-1/4\pi)\ln(f)$ term (see Appendix).
In the Hartle-Hawking state $T=1/4\pi$ and, hence, total
${\langle\varphi^2\rangle}_{\rm ren}$ is finite, as it should be.

Thus the uniform approximation solves the problem of finiteness of
$\langle\varphi^2\rangle$ at the horizon by accurately taking into
account of the zero mode contribution. Nevertheless
finiteness itself on the horizon does not guarantee the correct
value. So, we should compare our approximation with numerical
calculations to get an idea how precise it is. The most important
contribution comes from zero modes, this is why we present here
plots for comparison of only their contribution. For higher modes
the WKB approximation works very well (see, e.g.,
\cite{AnHiSa:95,Su}
\begin{figure}[h]
\centerline{
\includegraphics{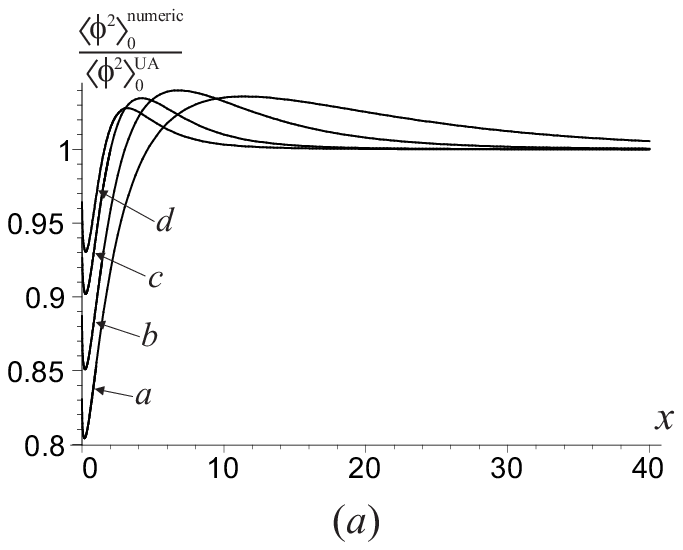}\quad
\includegraphics{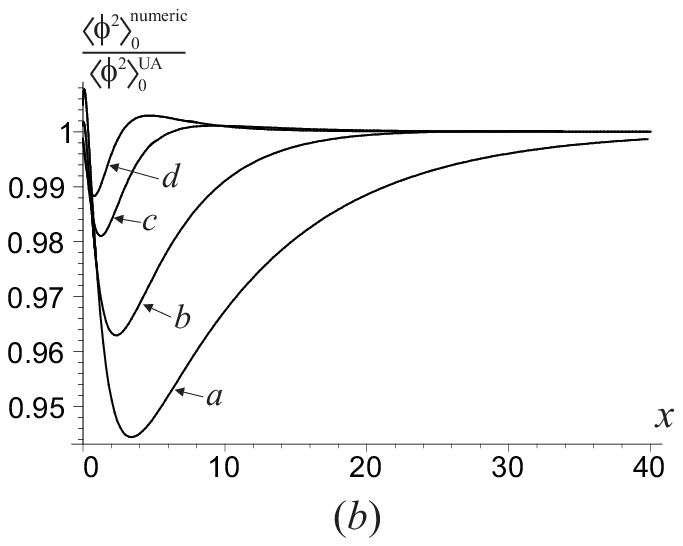}}
\caption[fig.3]{The graphs demonstrate the ratio of exact
numerical calculations and the uniform approximation for
${\langle\varphi^2\rangle}_0$. The figures (a) and (b) correspond
to $\xi=0$ and $\xi=1/6$, and the curves $a,b,c,d$ correspond to
$m=0.05,0.1,0.2,0.3$, respectively.} \label{numerics}
\end{figure}
One can see that the larger mass the more accurate becomes  the
uniform approximation. For $m \ge 1$ and $\xi\ge 1/6$ the uniform
approximation gives practically exact value on the horizon and at infinity
and the maximum deviation from exact function is less than 0.25\%.
For other parameters it is still very precise. This you can see in
Fig.(\ref{numerics}).

Let us summarize the obtained results. We demonstrated that the WKB
approximation traditionally used for obtaining different analytical
approximations for local observables breaks down for low frequency modes.
In the general case this results in the logarithmic divergences of
these observables at the black hole horizon. The adopted
approximations can be improved if one uses a more accurate
expression for the low frequency modes. We propose a method, which we
call a uniform approximation, which not only automatically gives a
finite value for local observables, but also provides a good
approximation of the observables uniformly in the interval from the
horizon to infinity. Our concrete calculations were done for $\langle
\varphi^2\rangle$ in 2D black hole spacetime, but the proposed method
can be used for other local observables and for higher dimensional
black holes.

\section*{Acknowledgment}

We would like to thank the Killam Trust for its support.
S.S. was also supported by the Russian Foundation for Basic Research
grant No 02-02-17177. S.S. is grateful to Valery Frolov and
Andrei Zelnikov for hospitality.

\begin{appendix}
\section{Remarks on the $\log$-divergence of
${\langle\varphi^2\rangle}$ and $\Tmn$ near the horizon}\label{A}

In this appendix we discuss the problem which accompanies many of
approximate methods [see, for example
\cite{Page:82,AnHiSa:95,FZ:87}] derived for calculating
expectation values, such as ${\langle\varphi^2\rangle}$ and
$\Tmn$, on the black hole background. The essence of the problem
is that the approximate expressions for these values turn out to
be proportional to $\ln f$ and hence diverge logarithmically near
the horizon where $f\to0$. It is important to stress that such the
log-divergence takes place for both vacuum and thermal expectation
values with any temperature, including the Hartle-Hawking state
with the Hawking temperature. We will call this problem as {\em
the problem of\, $\log$-divergence}.

As we will see the problem of $\log$-divergence is tightly
connected with accurate consideration and taking into account a
contribution of zero modes into expectation values near the
horizon. Really, as has been shown above the radial modes of a scalar
field with arbitrary mass and coupling near the horizon have the
asymptotical form \Ref{near0}. Then, the radial Green functions
${\cal G}_n=s_n^{(1)}s_n^{(2)}$ near the horizon are
\beq
{\cal G}_{n=0}=-\ln y,
\eeq
\beq
{\cal G}_{n>0}=\frac1{2\omega_n}=\frac{1}{4\pi nT},
\eeq
and the unrenormalized expression for ${\langle\varphi^2\rangle}$ reads
\beq
\langle\varphi^2\rangle_{\rm unren}=-T\ln y
+\frac{1}{2\pi}\sum_{n=1}^{\infty} \frac{\cos(2\pi
nT\epsilon)}{n},
\eeq
where $\epsilon=\tau-\tau'$. Using the formula
\beq\label{sum_n}
\sum_{n=1}^\infty\frac{\cos(n\kappa\epsilon)}{n}=
\frac12\ln\left[\frac1{2\left(1-\cos(\kappa\epsilon)\right)}\right]=
-\frac12\ln\kappa^2\epsilon^2+\calO(\epsilon^2),
\eeq
we obtain the asymptotical form for $\langle\varphi^2\rangle_{\rm
unren}$ near the horizon:
\beq
\langle\varphi^2\rangle_{\rm unren}=-T\ln y
-\frac1{4\pi}\ln\kappa^2\epsilon^2+\calO(\epsilon^2),
\eeq
where $\kappa=2\pi T$. To calculate a renormalized expression for
${\langle\varphi^2\rangle}$ one should subtract from $\langle\varphi^2\rangle_{\rm
unren}$ the DeWitt-Schwinger counerterms \Ref{phiDS} and go to the
limit $\epsilon\to0$: ${\langle\varphi^2\rangle}_{\rm ren}=\lim_{\epsilon\to0}
\left({\langle\varphi^2\rangle}_{\rm unren}-{\langle\varphi^2\rangle}_{\rm DS}\right)$. Note that
${\langle\varphi^2\rangle}_{\rm DS}\sim-(4\pi)^{-1}\ln\sigma$, and $\sigma\sim
f\epsilon^2$ in the limit $\epsilon\to0$. Hence
$$
{\langle\varphi^2\rangle}_{\rm ren}\sim-T\ln y+(4\pi)^{-1}\ln f+\calO(y^0).
$$
Taking into account that $f\approx y$ at $y\to 0$ we rediscover the
known result that ${\langle\varphi^2\rangle}_{\rm ren}$ is {\em regular} near the
horizon provided the temperature has the Hawking value,
$T=(4\pi)^{-1}$, otherwise ${\langle\varphi^2\rangle}_{\rm ren}$ diverges as $\ln y$.

It is worth noting that the term $(4\pi)^{-1}\ln f$ coming from
the DeWitt-Schwinger counterterms ${\langle\varphi^2\rangle}_{\rm DS}$ is compensated
by the term $T\ln y$ coming from the expression for the $n=0$
radial Green function ${\cal G}_0$. Thus we see that the zero
modes play an important role in the problem of $\log$-divergence.

To illustrate this statement we consider the approximation derived
by Anderson, Hiscock and Samuel \cite{AnHiSa:95}. They obtained the
approximate expressions for ${\langle\varphi^2\rangle}$ and $\Tmn$ [see Eqs. (4.2),
(4.3), (4.4)] which are proportional to $\ln f$ and hence diverge
logarithmically near the horizon where $f\to0$. The cause of this
divergence can be easily explained now. Examining the procedure
derived by Anderson, Hiscock and Samuel for computing ${\langle\varphi^2\rangle}$ and
$\Tmn$ for the nonzero temperature case one faces with a necessity
to calculate the mode sums over $n$. Since the authors use for
this aim an expansion in inverse powers of $n$, they encounter the
problem of including into the calculation scheme a contribution of
the $n=0$ modes. To resolve this problem the authors impose a
lower limit cutoff in the mode sums, which for the nonzero
temperature case means merely discarding the $n=0$ contribution
from the summation procedure (see a discussion in the section IV
and the appendix E). But as was shown above, taking into account
the $n=0$ modes is {\em necessary} for a compensation of the
$\log$-contribution coming from the counterterms ${\langle\varphi^2\rangle}_{\rm DS}$
and $\Tmn_{\rm DS}$; otherwise, one gets the problem of
log-divergence.

\end{appendix}


\end{document}